\RequirePackage[running]{lineno}
\documentclass[aps,prl,nofootinbib,twocolumn,showpacs,superscriptaddress,letterpaper,amsmath,amsfonts,amssymb, preprintnumbers]{revtex4-1}
\pdfoutput=1 

\usepackage[bookmarks=false]{hyperref}

\usepackage{graphicx}
\usepackage{dcolumn}
\usepackage{bm}
\usepackage{xspace}
\usepackage{siunitx}
\usepackage[italic]{hepnames}
\usepackage{xcolor}
\usepackage{orcidlink}

\usepackage{rotating}
\usepackage{booktabs}

\DeclareSIUnit{\barn}{b}

\let\oldalign\align
\let\oldendalign\endalign

\renewenvironment{align}
  {\linenomathNonumbers\oldalign}
  {\oldendalign\endlinenomath}

\newcommand{\fb}[1]{{\color{pink}#1}}

\newcolumntype{d}[1]{D{.}{.}{#1}}

\def\lumi{\SI{35.4}{\per\femto\barn}}
\def\lumiError{$(35.4 \pm 0.8)$ \SI{}{\per\femto\barn}} 

\def\obssign{\ensuremath{16}}
\def\nuexp{\ensuremath{ n_\nu^{\mathrm{exp}}  = 151 \pm 41}}

\makeatletter
\g@addto@macro\bfseries{\boldmath}
\makeatother
\allowdisplaybreaks[2]

\providecommand{\papersection[1]}{\paragraph{#1}}

\begin{document}


\title{First Direct Observation of Collider Neutrinos with FASER at the LHC \\
\vspace*{0.08in} 
{\normalsize FASER Collaboration} 
}

\author{\vspace{-0.12in} Henso Abreu\,\orcidlink{0000-0002-1599-2896}} 
\affiliation{Department of Physics and Astronomy, Technion---Israel Institute of Technology, Haifa 32000, Israel}

\author{John Anders\,\orcidlink{0000-0002-1846-0262}} 
\affiliation{CERN, CH-1211 Geneva 23, Switzerland}

\author{Claire Antel\,\orcidlink{0000-0001-9683-0890}} 
\affiliation{D\'epartement de Physique Nucl\'eaire et Corpusculaire, University of Geneva, CH-1211 Geneva 4, Switzerland}

\author{Akitaka Ariga\,\orcidlink{0000-0002-6832-2466}} 
\affiliation{Albert Einstein Center for Fundamental Physics, Laboratory for High Energy Physics, University of Bern, Sidlerstrasse 5, CH-3012 Bern, Switzerland}
\affiliation{Department of Physics, Chiba University, 1-33 Yayoi-cho Inage-ku, 263-8522 Chiba, Japan}

\author{Tomoko Ariga\,\orcidlink{0000-0001-9880-3562}} 
\affiliation{Kyushu University, Nishi-ku, 819-0395 Fukuoka, Japan}

\author{Jeremy Atkinson\,\orcidlink{0009-0003-3287-2196}}	
\affiliation{Albert Einstein Center for Fundamental Physics, Laboratory for High Energy Physics, University of Bern, Sidlerstrasse 5, CH-3012 Bern, Switzerland}

\author{Florian~U.~Bernlochner\,\orcidlink{0000-0001-8153-2719}} 
\affiliation{Universit\"at Bonn, Regina-Pacis-Weg 3, D-53113 Bonn, Germany}

\author{Tobias Blesgen\,\orcidlink{0009-0007-2013-1031}} 
\affiliation{Universit\"at Bonn, Regina-Pacis-Weg 3, D-53113 Bonn, Germany}

\author{Tobias Boeckh\,\orcidlink{0009-0000-7721-2114}} 
\affiliation{Universit\"at Bonn, Regina-Pacis-Weg 3, D-53113 Bonn, Germany}

\author{Jamie Boyd\,\orcidlink{0000-0001-7360-0726}} 
\affiliation{CERN, CH-1211 Geneva 23, Switzerland}

\author{Lydia Brenner\,\orcidlink{0000-0001-5350-7081}} 
\affiliation{Nikhef National Institute for Subatomic Physics, Science Park 105, 1098 XG Amsterdam, Netherlands}

\author{Franck Cadoux} 
\affiliation{D\'epartement de Physique Nucl\'eaire et Corpusculaire, University of Geneva, CH-1211 Geneva 4, Switzerland}

\author{David~W.~Casper\,\orcidlink{0000-0002-7618-1683}} 
\affiliation{Department of Physics and Astronomy, University of California, Irvine, CA 92697-4575, USA}

\author{Charlotte Cavanagh\,\orcidlink{0009-0001-1146-5247}} 
\affiliation{University of Liverpool, Liverpool L69 3BX, United Kingdom}

\author{Xin Chen\,\orcidlink{0000-0003-4027-3305}} 
\affiliation{Department of Physics, Tsinghua University, Beijing, China}

\author{Andrea Coccaro\,\orcidlink{0000-0003-2368-4559}} 
\affiliation{INFN Sezione di Genova, Via Dodecaneso, 33--16146, Genova, Italy}

\author{Ansh Desai\,\orcidlink{0000-0002-5447-8304}} 
\affiliation{University of Oregon, Eugene, OR 97403, USA}

\author{Sergey Dmitrievsky\,\orcidlink{0000-0003-4247-8697}} 
\affiliation{Affiliated with an international laboratory covered by a cooperation agreement with CERN.}

\author{Monica D’Onofrio\,\orcidlink{0000-0003-2408-5099}} 
\affiliation{University of Liverpool, Liverpool L69 3BX, United Kingdom}

\author{Yannick Favre}
\affiliation{D\'epartement de Physique Nucl\'eaire et Corpusculaire, University of Geneva, CH-1211 Geneva 4, Switzerland}

\author{Deion Fellers\,\orcidlink{0000-0002-0731-9562}} 
\affiliation{University of Oregon, Eugene, OR 97403, USA}

\author{Jonathan~L.~Feng\,\orcidlink{0000-0002-7713-2138}} 
\affiliation{Department of Physics and Astronomy, 
University of California, Irvine, CA 92697-4575, USA}

\author{Carlo Alberto Fenoglio\,\orcidlink{0009-0007-7567-8763}}	
\affiliation{D\'epartement de Physique Nucl\'eaire et Corpusculaire, University of Geneva, CH-1211 Geneva 4, Switzerland}

\author{Didier Ferrere\,\orcidlink{0000-0002-5687-9240}} 
\affiliation{D\'epartement de Physique Nucl\'eaire et Corpusculaire, University of Geneva, CH-1211 Geneva 4, Switzerland}

\author{Stephen Gibson\,\orcidlink{0000-0002-1236-9249}} 
\affiliation{Royal Holloway, University of London, Egham, TW20 0EX, United Kingdom}

\author{Sergio Gonzalez-Sevilla\,\orcidlink{0000-0003-4458-9403}} 
\affiliation{D\'epartement de Physique Nucl\'eaire et Corpusculaire, University of Geneva, CH-1211 Geneva 4, Switzerland}

\author{Yuri Gornushkin\,\orcidlink{0000-0003-3524-4032}} 
\affiliation{Affiliated with an international laboratory covered by a cooperation agreement with CERN.}

\author{Carl Gwilliam\,\orcidlink{0000-0002-9401-5304}} 
\affiliation{University of Liverpool, Liverpool L69 3BX, United Kingdom}

\author{Daiki Hayakawa\,\orcidlink{0000-0003-4253-4484}} 
\affiliation{Department of Physics, Chiba University, 1-33 Yayoi-cho Inage-ku, 263-8522 Chiba, Japan}

\author{Shih-Chieh Hsu\,\orcidlink{0000-0001-6214-8500}} 
\affiliation{Department of Physics, University of Washington, PO Box 351560, Seattle, WA 98195-1460, USA}

\author{Zhen Hu\,\orcidlink{0000-0001-8209-4343}} 
\affiliation{Department of Physics, Tsinghua University, Beijing, China}

\author{Giuseppe Iacobucci\,\orcidlink{0000-0001-9965-5442}} 
\affiliation{D\'epartement de Physique Nucl\'eaire et Corpusculaire, University of Geneva, CH-1211 Geneva 4, Switzerland}

\author{Tomohiro Inada\,\orcidlink{0000-0002-6923-9314}} 
\affiliation{Department of Physics, Tsinghua University, Beijing, China}

\author{Sune Jakobsen\,\orcidlink{0000-0002-6564-040X}} 
\affiliation{CERN, CH-1211 Geneva 23, Switzerland}

\author{Hans Joos\,\orcidlink{0000-0003-4313-4255}}	
\affiliation{CERN, CH-1211 Geneva 23, Switzerland}
\affiliation{II.~Physikalisches Institut, Universität Göttingen, Göttingen, Germany}

\author{Enrique Kajomovitz\,\orcidlink{0000-0002-8464-1790}} 
\affiliation{Department of Physics and Astronomy, Technion---Israel Institute of Technology, Haifa 32000, Israel}

\author{Hiroaki Kawahara\,\orcidlink{0009-0007-5657-9954}} 
\affiliation{Kyushu University, Nishi-ku, 819-0395 Fukuoka, Japan}

\author{Alex Keyken}	
\affiliation{Royal Holloway, University of London, Egham, TW20 0EX, United Kingdom}

\author{Felix Kling\,\orcidlink{0000-0002-3100-6144}} 
\affiliation{Deutsches Elektronen-Synchrotron DESY, Notkestr. 85, 22607 Hamburg, Germany}

\author{Daniela Köck\,\orcidlink{0000-0002-9090-5502}} 
\affiliation{University of Oregon, Eugene, OR 97403, USA}

\author{Umut Kose\,\orcidlink{0000-0001-5380-9354}} 
\affiliation{CERN, CH-1211 Geneva 23, Switzerland}

\author{Rafaella Kotitsa\,\orcidlink{0000-0002-7886-2685}} 
\affiliation{CERN, CH-1211 Geneva 23, Switzerland}

\author{Susanne Kuehn\,\orcidlink{0000-0001-5270-0920}} 
\affiliation{CERN, CH-1211 Geneva 23, Switzerland}

\author{Helena Lefebvre\,\orcidlink{0000-0002-7394-2408}} 
\affiliation{Royal Holloway, University of London, Egham, TW20 0EX, United Kingdom}

\author{Lorne Levinson\,\orcidlink{0000-0003-4679-0485}} 
\affiliation{Department of Particle Physics and Astrophysics, Weizmann Institute of Science, Rehovot 76100, Israel}

\author{Ke Li\,\orcidlink{0000-0002-2545-0329}} 
\affiliation{Department of Physics, University of Washington, PO Box 351560, Seattle, WA 98195-1460, USA}

\author{Jinfeng Liu}
\affiliation{Department of Physics, Tsinghua University, Beijing, China}

\author{Jack MacDonald\,\orcidlink{0000-0002-3150-3124}}	
\affiliation{Institut f\"ur Physik, Universität Mainz, Mainz, Germany}

\author{Chiara Magliocca\,\orcidlink{0009-0009-4927-9253}} 
\affiliation{D\'epartement de Physique Nucl\'eaire et Corpusculaire, University of Geneva, CH-1211 Geneva 4, Switzerland}

\author{Fulvio Martinelli\,\orcidlink{0000-0003-4221-5862}} 
\affiliation{D\'epartement de Physique Nucl\'eaire et Corpusculaire, University of Geneva, CH-1211 Geneva 4, Switzerland}

\author{Josh McFayden\,\orcidlink{0000-0001-9273-2564}} 
\affiliation{Department of Physics \& Astronomy, University of Sussex, Sussex House, Falmer, Brighton, BN1 9RH, United Kingdom}

\author{Matteo Milanesio\,\orcidlink{0000-0001-8778-9638}} 
\affiliation{D\'epartement de Physique Nucl\'eaire et Corpusculaire, University of Geneva, CH-1211 Geneva 4, Switzerland}

\author{Dimitar Mladenov\,\orcidlink{0000-0002-0692-0495}} 
\affiliation{CERN, CH-1211 Geneva 23, Switzerland}

\author{Théo Moretti\,\orcidlink{0000-0001-7065-1923}} 
\affiliation{D\'epartement de Physique Nucl\'eaire et Corpusculaire, University of Geneva, CH-1211 Geneva 4, Switzerland}

\author{Magdalena Munker\,\orcidlink{0000-0003-2775-3291}} 
\affiliation{D\'epartement de Physique Nucl\'eaire et Corpusculaire, University of Geneva, CH-1211 Geneva 4, Switzerland}

\author{Mitsuhiro Nakamura}
\affiliation{Nagoya University, Furo-cho, Chikusa-ku, Nagoya 464-8602, Japan}

\author{Toshiyuki Nakano}
\affiliation{Nagoya University, Furo-cho, Chikusa-ku, Nagoya 464-8602, Japan}

\author{Marzio Nessi\,\orcidlink{0000-0001-7316-0118}} 
\affiliation{D\'epartement de Physique Nucl\'eaire et Corpusculaire,  University of Geneva, CH-1211 Geneva 4, Switzerland}
\affiliation{CERN, CH-1211 Geneva 23, Switzerland}

\author{Friedemann Neuhaus\,\orcidlink{0000-0002-3819-2453}} 
\affiliation{Institut f\"ur Physik, Universität Mainz, Mainz, Germany}

\author{Laurie Nevay\,\orcidlink{0000-0001-7225-9327}} 
\affiliation{CERN, CH-1211 Geneva 23, Switzerland}
\affiliation{Royal Holloway, University of London, Egham, TW20 0EX, United Kingdom}

\author{Hidetoshi Otono\,\orcidlink{0000-0003-0760-5988}} 
\affiliation{Kyushu University, Nishi-ku, 819-0395 Fukuoka, Japan}

\author{Hao Pang\,\orcidlink{0000-0002-1946-1769}} 
\affiliation{Department of Physics, Tsinghua University, Beijing, China}

\author{Lorenzo Paolozzi\,\orcidlink{0000-0002-9281-1972}} 
\affiliation{D\'epartement de Physique Nucl\'eaire et Corpusculaire, University of Geneva, CH-1211 Geneva 4, Switzerland}
\affiliation{CERN, CH-1211 Geneva 23, Switzerland}

\author{Brian Petersen\,\orcidlink{0000-0002-7380-6123}} 
\affiliation{CERN, CH-1211 Geneva 23, Switzerland}

\author{Francesco Pietropaolo}
\affiliation{CERN, CH-1211 Geneva 23, Switzerland}

\author{Markus Prim\,\orcidlink{0000-0002-1407-7450}} 
\affiliation{Universit\"at Bonn, Regina-Pacis-Weg 3, D-53113 Bonn, Germany}

\author{Michaela Queitsch-Maitland\,\orcidlink{0000-0003-4643-515X}} 
\affiliation{University of Manchester, School of Physics and Astronomy, Schuster Building, Oxford Rd, Manchester M13 9PL, United Kingdom}

\author{Filippo Resnati\,\orcidlink{0000-0003-1434-6435}} 
\affiliation{CERN, CH-1211 Geneva 23, Switzerland}

\author{Hiroki Rokujo}
\affiliation{Nagoya University, Furo-cho, Chikusa-ku, Nagoya 464-8602, Japan}

\author{Elisa Ruiz-Choliz\,\orcidlink{0000-0002-2417-7121}} 
\affiliation{Institut f\"ur Physik, Universität Mainz, Mainz, Germany}

\author{Jorge Sabater-Iglesias\,\orcidlink{0000-0003-2328-1952}} 
\affiliation{D\'epartement de Physique Nucl\'eaire et Corpusculaire, University of Geneva, CH-1211 Geneva 4, Switzerland}

\author{Osamu Sato\,\orcidlink{0000-0002-6307-7019}} 
\affiliation{Nagoya University, Furo-cho, Chikusa-ku, Nagoya 464-8602, Japan}

\author{Paola Scampoli\,\orcidlink{0000-0001-7500-2535}} 
\affiliation{Albert Einstein Center for Fundamental Physics, Laboratory for High Energy Physics, University of Bern, Sidlerstrasse 5, CH-3012 Bern, Switzerland}
\affiliation{Dipartimento di Fisica ``Ettore Pancini'', Universit\`a di Napoli Federico II, Complesso Universitario di Monte S. Angelo, I-80126 Napoli, Italy}

\author{Kristof Schmieden\,\orcidlink{0000-0003-1978-4928}} 
\affiliation{Institut f\"ur Physik, Universität Mainz, Mainz, Germany}

\author{Matthias Schott\,\orcidlink{0000-0002-4235-7265}} 
\affiliation{Institut f\"ur Physik, Universität Mainz, Mainz, Germany}

\author{Anna Sfyrla\,\orcidlink{0000-0002-3003-9905}} 
\affiliation{D\'epartement de Physique Nucl\'eaire et Corpusculaire, University of Geneva, CH-1211 Geneva 4, Switzerland}

\author{Savannah Shively\,\orcidlink{0000-0002-4691-3767}} 
\affiliation{Department of Physics and Astronomy, University of California, Irvine, CA 92697-4575, USA}

\author{Yosuke Takubo\,\orcidlink{0000-0002-3143-8510}} 
\affiliation{Institute of Particle and Nuclear Studies, KEK, Oho 1-1, Tsukuba, Ibaraki 305-0801, Japan}

\author{Noshin Tarannum\,\orcidlink{0000-0002-3246-2686}} 
\affiliation{D\'epartement de Physique Nucl\'eaire et Corpusculaire, University of Geneva, CH-1211 Geneva 4, Switzerland}

\author{Ondrej Theiner,\orcidlink{0000-0002-6558-7311}} 
\affiliation{D\'epartement de Physique Nucl\'eaire et Corpusculaire, University of Geneva, CH-1211 Geneva 4, Switzerland}

\author{Eric Torrence\,\orcidlink{0000-0003-2911-8910}} 
\affiliation{University of Oregon, Eugene, OR 97403, USA}

\author{Serhan Tufanli}
\affiliation{CERN, CH-1211 Geneva 23, Switzerland}

\author{Svetlana Vasina\,\orcidlink{0000-0003-2775-5721}} 
\affiliation{Affiliated with an international laboratory covered by a cooperation agreement with CERN.}

\author{Benedikt Vormwald\,\orcidlink{0000-0003-2607-7287}} 
\affiliation{CERN, CH-1211 Geneva 23, Switzerland}

\author{Di Wang\,\orcidlink{0000-0002-0050-612X}} 
\affiliation{Department of Physics, Tsinghua University, Beijing, China}

\author{Eli Welch\,\orcidlink{0000-0001-6336-2912}} 
\affiliation{Department of Physics and Astronomy, University of California, Irvine, CA 92697-4575, USA}

\author{Stefano Zambito\,\orcidlink{0000-0002-4499-2545}} 
\affiliation{D\'epartement de Physique Nucl\'eaire et Corpusculaire, University of Geneva, CH-1211 Geneva 4, Switzerland}

\begin{abstract}
We report the first direct observation of neutrino interactions at a particle collider experiment. Neutrino candidate events are identified in a \SI{13.6}{\TeV} center-of-mass energy $pp$ collision data set of \lumi using the active electronic components of the FASER detector at the Large Hadron Collider. The candidates are required to have a track propagating through the entire length of the FASER detector and be consistent with a muon neutrino charged-current interaction. We infer $153^{+12}_{-13}$ neutrino interactions with a significance of \obssign\ standard deviations above the background-only hypothesis. These events are consistent with the characteristics expected from neutrino interactions in terms of secondary particle production and spatial distribution, and they imply the observation of both neutrinos and anti-neutrinos with an incident neutrino energy of significantly above 200 GeV.
\end{abstract}

\preprint{CERN-EP-2023-056}

\date{March 24, 2023}

\maketitle

\onecolumngrid

\begin{center}
 © 2023 CERN for the benefit of the FASER Collaboration.
Reproduction of this article or parts of it is allowed as specified in the CC-BY-4.0 license.
\end{center}

\twocolumngrid

\papersection{Introduction}

Since their discovery at a nuclear reactor in 1956~\cite{Cowan:1992xc}, neutrinos have been detected from a variety of sources:~fixed target experiments~\cite{Danby:1962nd}, cosmic ray interactions in the atmosphere~\cite{Achar:1965ova, Reines:1965qk, Fukuda:1998mi}, the Sun~\cite{Davis:1968cp, Cleveland:1998nv}, the Earth~\cite{Araki:2005qa}, supernovae~\cite{Hirata:1987hu, Bionta:1987qt}, and astrophysical bodies outside our galaxy~\cite{Aartsen:2014gkd}. With each new source has come new insights, with important implications for many fields, from particle physics to geophysics to astrophysics and cosmology.  

Until now, however, no neutrino produced at a particle collider has ever been directly detected. Colliders copiously produce both neutrinos and anti-neutrinos of all flavors, and they do so in a range of very high energies where neutrino interactions have not yet been observed.  Nevertheless, collider neutrinos have escaped detection, because they interact extremely weakly, and the highest energy neutrinos, which have the largest probability of interacting, are dominantly produced in the forward region, parallel to the beamline~\cite{DeRujula:1984pg, DeRujula:1984ns, Vannucci:253670, DeRujula:1992sn, Park:2011gh, Feng:2017uoz, FASER:2019dxq}, where collider detectors typically have uninstrumented regions to allow the entry and exit of the colliding particle beams. In 2021, the FASER Collaboration identified the first collider neutrino candidates~\cite{FASER:2021mtu} using a \SI{29}{\kg} pilot detector,
highlighting the potential of discovering collider neutrinos in the forward region of the Large Hadron Collider (LHC) collisions. In addition to FASER, the SND@LHC experiment is expected to observe and study neutrinos produced in the LHC collisions~\cite{Beni:2019gxv,SHiP:2020sos,SNDLHC:2022ihg} and recently reported preliminary findings~\cite{BLAH}. The observation of collider neutrino interactions will have wide ranging implications for the study of neutrino properties, QCD, astroparticle physics, and searches for physics beyond the Standard Model~\cite{Feng:2022inv}.

This letter reports the first direct observation of neutrinos produced at a particle collider by analyzing \lumi of proton-proton ($pp$) collision data from Run 3 of the LHC at a center-of-mass energy of \SI{13.6}{\TeV}. Neutrinos of all flavors are produced in the decays of light and heavy hadrons as a high-intensity beam 
along the collision axis. In this paper we focus on the charged-current (CC) interactions of $\nu_\mu$ and $\overline \nu_\mu$; in the following, charge conjugation and natural units are implied throughout. The chosen analysis strategy is designed to be independent of the simulation of the detector response and therefore does not measure the neutrino interaction cross section, but rather the significance of the observed number of neutrino candidate events over the non-neutrino background. 
In addition to being the first collider neutrinos ever observed, the neutrinos detected here are expected to be the most energetic ever detected from a human source, with energies in the unexplored range \SI{360}{\GeV}-\SI{6.3}{\TeV} between fixed target measurements~\cite{Zyla:2020zbs} and astroparticle data~\cite{Aartsen:2017kpd}.

\papersection{The FASER Detector}

 \begin{figure*}
	\centering
	\includegraphics[width=1.0\linewidth]{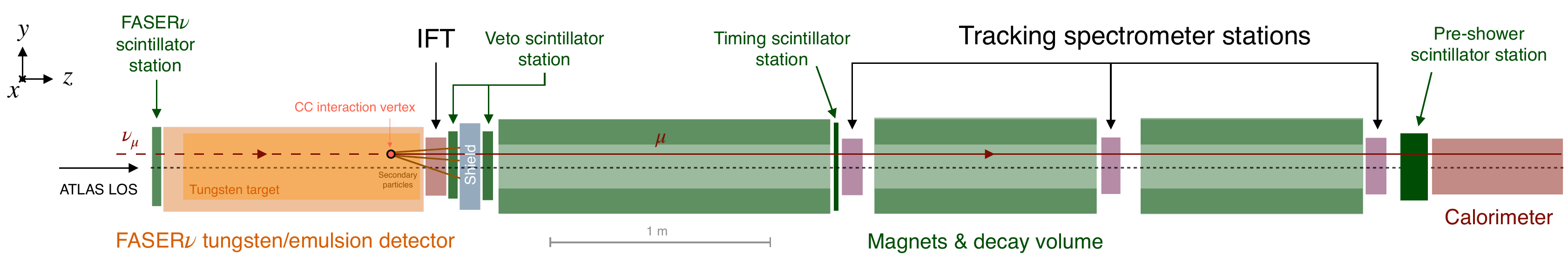}   
\vspace*{-0.15in}
\caption{ 
	Schematic side view of the FASER detector with a muon neutrino undergoing a CC interaction in the emulsion-tungsten target. 
	}
\label{fig:faser} 	
\end{figure*}

FASER~\cite{FASER:2022hcn, Feng:2017uoz, FASER:2018ceo, FASER:2018bac}, the ForwArd Search ExpeRiment, is an apparatus dedicated to searching for light, extremely weakly-interacting particles and studying neutrinos. A detailed description can be found in Ref.~\cite{FASER:2022hcn}. The experiment is located in the TI12 tunnel, which connects the Super Proton Synchrotron (SPS) and LHC tunnels, approximately \SI{480}{\meter} downstream of the ATLAS interaction point (IP) and aligned with the collision axis line-of-sight (LOS). Charged particles produced in the forward direction at the ATLAS IP are deflected by LHC magnets, and FASER is also shielded from the ATLAS IP by about \SI{100}{\meter} of rock and concrete.  FASER's location therefore ensures that a high-intensity beam of neutrinos traverses the detector, while backgrounds are highly suppressed.

The FASER detector is partially immersed in a magnetic field and consists of a passive tungsten-emulsion neutrino detector (FASER$\nu$), two scintillator-based veto systems, additional scintillators for triggering, a tracking spectrometer, a pre-shower scintillator station, and an electromagnetic calorimeter. For the current analysis, the most important components are the veto systems, the tracking spectrometer, and the tungsten target of the FASER$\nu$ detector. A schematic of the FASER detector is depicted in Figure~\ref{fig:faser}. The trigger and data acquisition system of FASER was designed to achieve high efficiency and reliability~\cite{FASER:2021cpr}. Neutrino candidate events are triggered by scintillator signals that exceed a preset threshold below that of a single minimum-ionizing particle (MIP), resulting in a typical trigger rate of 0.5-\SI{1.3}{\kilo\Hz}. 

FASER$\nu$ consists of 730 layers of \SI{1.1}{\milli\meter}-thick tungsten plates interleaved with emulsion films. With a width of \SI{25}{\centi\meter} and a height of \SI{30}{\centi\meter}, it has a total mass of 1.1 metric tons. Although the emulsion films provide excellent position and angular resolution to identify CC neutrino interactions, they are not used as the extraction, scanning and analysis is time intensive. Instead, the FASER$\nu$ detector is used as a target for CC neutrino interactions, and we rely on the active electronic detector components of FASER to identify suitable muon neutrino candidates~\cite{Arakawa:2022rmp}. 

The FASER scintillator stations are instrumental to identify suitable neutrino candidates and veto charged particles originating from the interaction point or from secondary interactions. The first veto system (FASER$\nu$ scintillator station) is located in front of the FASER$\nu$ emulsion detector. It is constructed from two modules of $\SI{30}{\centi\meter} \times \SI{35}{\centi\meter}$, \SI{2}{\centi\meter}-thick plastic scintillators, which are read out with photomultiplier tubes (PMTs). The second veto system (veto scintillator station) is located after the FASER$\nu$ emulsion detector and in front of the first magnet. It is built from three planes of the same plastic scintillators, arranged with a \SI{10}{\centi\meter}-thick lead block placed between the first and second planes. The lead acts as an additional target for neutrino interactions and to absorb or convert high-energy photons from muon bremsstrahlung. 

The tracking system consists of the interface tracking station (IFT) and the three tracking spectrometer stations~\cite{FASER:2021ljd}. Each tracking station is composed of three planes with eight ATLAS semiconductor tracker (SCT) barrel modules~\cite{Abdesselam:2006wt} per plane, arranged as two columns of four modules. Each SCT module consists of a double-layer of single-sided silicon microstrips with a \SI{40}{\milli\radian}  stereo angle and an \SI{80}{\um} strip pitch. To identify muons from CC interactions, only the tracking spectrometer stations are used, whereas the IFT's location after the tungsten-emulsion detector makes it ideal to study remnants and secondary particles of CC deep inelastic scattering neutrino interactions. A muon candidate traversing the full length of the spectrometer produces 18 silicon hits. Adjacent silicon hits in the tracking stations are combined into clusters. Between the three tracking spectrometer stations are two \SI{1}{\meter}-long dipole magnets with magnetic field of 0.57~T, with a similar \SI{1.5}{\meter}-long magnet in front of the spectrometer. The magnets have an aperture of 200 mm diameter, which defines the active transverse area of the detector, and bend charged particles in the vertical plane. In addition, signals from the timing scintillator station, located between the first and second magnet and in front of the first tracking station of the spectrometer, are used. The scintillator stations in combination with the tracking system are capable of reliably identifying incoming charged particles passing through the full length of FASER with inefficiencies smaller than $10^{-7}$, depending on the momentum and other requirements in the selection.

\papersection{Data Set and Simulated Samples}

For this analysis we use data from runs with stable beam conditions collected between July and November 2022, corresponding to a total luminosity of \lumiError~\cite{ATLAS:2022hro,ATLAS:2016fhk}  after data quality selection. A detailed description of the analysis is following; additional details are contained in Appendices A-E.

To study the detector response to neutrino interactions, we simulate $4.3 \times 10^4$ neutrino events corresponding to an integrated luminosity of approximately \SI{600}{\per\femto\barn}. The interaction with the tungsten-emulsion detector is simulated using the \texttt{GENIE} event generator~\cite{Genie2010,Genie2015}. The neutrino energy spectra and relative flavor composition are based on Ref.~\cite{Kling:2021gos}. To estimate the number of expected neutrino events, we adjust several of the assumptions of Ref.~\cite{Kling:2021gos}:~we correct the center-of-mass energy, beam crossing angle, and LOS alignment, and we use the average of the neutrino flux from the predicted light and heavy hadron production of \texttt{DPMJET}~\cite{Roesler:2000he, Fedynitch:2015kcn} and \texttt{SIBYLL}~\cite{Riehn:2019jet}. The difference between the two individual predictions and the average is 27\% and is assigned as an uncertainty. All interactions of particles traversing the FASER detector are simulated using \texttt{GEANT4}~\cite{GEANT4:2002zbu}. 

The main background to neutrino signatures originates from high-momentum muons. We use the energy and angular spectrum predicted by the FLUKA generator~\cite{Fluka2005,Fluka2015}, which includes a detailed description of the LHC machine elements and infrastructure, to simulate a sample of $2 \times 10^6$ muons for background studies. Two additional sources of backgrounds are relevant:~neutral hadrons produced by muon interactions in the concrete in front of the FASER detector and geometric backgrounds from charged particles missing the FASER$\nu$ scintillator. 

We use simulated samples to study the neutral hadron backgrounds. The contamination from geometric background events is studied using sidebands and extrapolated into the signal region using simulations. The backgrounds from cosmic rays and LHC beam background have been studied using events occurring when there are no collisions, and are found to be negligible.

    \papersection{Selection and Background Rejection}

We focus on identifying $\nu_\mu$ and $\overline \nu_\mu$ CC interactions produced in the tungsten-emulsion detector. Such interactions will produce a high-momentum $\mu$ that can be reconstructed in the three stations of the FASER tracking spectrometer. In addition, we expect increased activity in the veto and timing scintillator stations and in the IFT tracking station from secondary particles produced in the CC interaction. To avoid unconscious bias, a blind analysis was carried out where the event selection, background estimations, and systematic uncertainties were fixed prior to looking at data in the signal-enhanced region.

We select events triggered by any of the scintillators downstream of FASER$\nu$. To discard signals from beam backgrounds and cosmic muons, we further require a timing stamp consistent with a colliding bunch crossing identifier. We use the FASER$\nu$ scintillator to identify backgrounds from muons or other charged particles entering the FASER detector and reject events that deposit a charge of more than \SI{40}{\pico\coulomb} in the PMTs. Such a charge deposition would be consistent with the presence of one or several MIPs. We only look for CC interactions that produce a muon that traverses the entire length of the FASER detector. The signals in the scintillators downstream of the lead wall in the veto system, and in the calorimeter, are therefore required to be compatible with those of a MIP. With the three tracker stations we reconstruct events with exactly one track and require more than 11 silicon hits in the tracking stations.  The reconstructed tracks are required to have a reasonable track fit quality, and we require the reconstructed track momentum to fulfill $p_\mu > \SI{100}{GeV}$. To reject charged particles, whose trajectory geometrically missed the FASER$\nu$ scintillator station, we extrapolate the reconstructed track from the spectrometer back to the IFT and FASER$\nu$ scintillator. The track's extrapolation to the IFT must lie within \SI{95}{\milli\meter} of the detector's central axis, and its extrapolation to the FASER$\nu$ scintillator must be at a distance of \mbox{$r_{\mathrm{veto}\,\nu} <$ \SI{120}{\milli\meter}} from the FASER$\nu$ scintillator center. 

\papersection{Neutral Hadron and Geometric Backgrounds}

To estimate the number of neutral hadrons that reach FASER, we simulate \SI{2.1e9}\ $\mu$ events based on the \texttt{FLUKA} energy spectrum, and use \texttt{GEANT4} to propagate through the last \SI{8}{m} of rock in front of FASER. From this sample we determine the number of neutral hadrons with a momentum larger than \SI{100}{GeV} that reach the detector. The selection efficiency is evaluated with an additional sample of neutral kaons and neutrons with momenta larger than \SI{100}{GeV} in front of the FASER$\nu$ emulsion detector. Most simulated hadrons are absorbed in the tungsten or do not produce a charged track with sufficient momentum to pass the signal selection and only a small fraction of the simulated hadrons pass all selection steps. From this we estimate the total neutral hadron background to be $n_{\mathrm{had}} = 0.11 \pm 0.06$, with the uncertainty denoting the statistical error. Further simulation studies show that in most cases the parent muon enters the detector along with the neutral hadron. Such events would be rejected by the FASER$\nu$ veto scintillator. The estimate assumes that all neutral hadron events are not already vetoed by the accompanying muon, and is therefore a conservative estimate of this background contribution.

To estimate the geometric background contribution, we count the number of background events $n_{\mathrm{geo}}$ using a sideband and apply a scaling to the signal region of $f_{\mathrm{geo}}$, which is extracted from simulated samples. The sideband is defined to enhance the contribution of muons that miss the FASER$\nu$ scintillator station, but may be able to produce a track in the spectrometer, which passes the selection by scattering in the tungsten and/or bending in the magnetic field. We modify the event selection outlined above: we require at most 8 IFT clusters, an extrapolated radius $r_{\mathrm{IFT}}$ of \SI{90}{\milli\meter} to \SI{95}{\milli\meter} with respect to the IFT center, and apply no selection on $r_{\mathrm{veto}\,\nu}$. None of the selected sideband events have a momentum larger than \SI{100}{GeV}. We thus extrapolate to the signal region by using a linear fit to the momentum distribution. We correct this estimate to account for the $r_{\mathrm{veto}\,\nu}$ selection by using the ratio of events with a radius smaller than \SI{120}{mm} over all sideband events in the fitted range. As we observe no events with $r_{\mathrm{veto}\,\nu} <\SI{120}{mm}$ in the fitted range, we use the 3$\sigma$ upper limit of the expectation value of a Poisson process for an observation of zero events of 5.9. With this we find $n_{\mathrm{geo}} = 0.01 \pm 0.23$ background events in the sideband, with the uncertainty denoting the statistical error. We extract a scaling factor between this sideband and the signal region from simulations, probing different momenta, angles, and position ranges, and use the resulting deviation from the nominal simulation scenario as an uncertainty. This results in a scaling factor of $f_{\mathrm{geo}}  = 7.9 \pm 2.4$ and a total geometric background estimate of $0.08 \pm 1.83$ events.

\papersection{Results}

 \begin{figure}[t]
		 \includegraphics[width=1.0\linewidth]{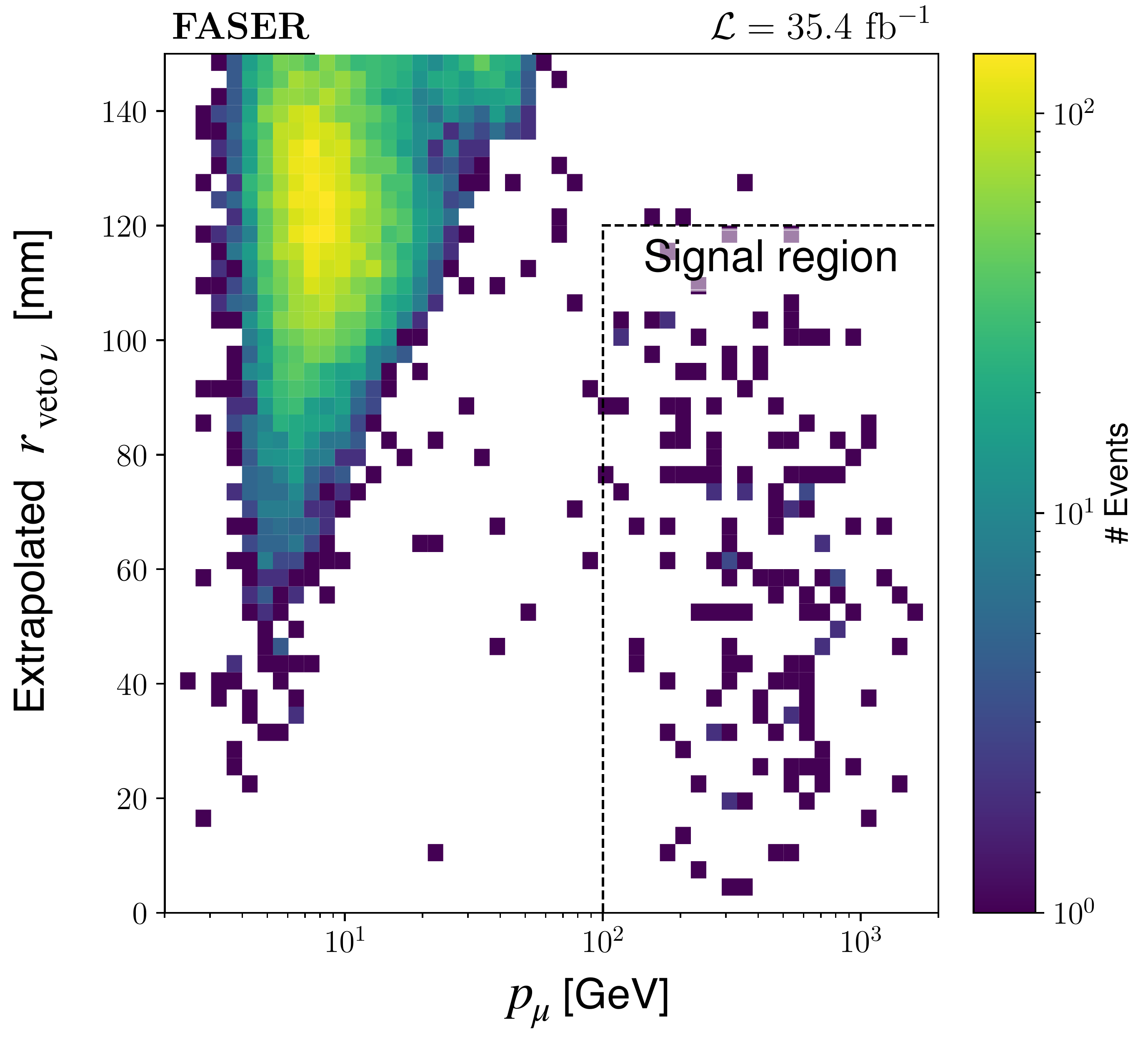}  
   \caption{ 
   		  The selected signal region in extrapolated radius $r_{\mathrm{veto}\, \nu}$ and reconstructed track momentum $p_\mu$ is depicted. The region with lower momenta and larger radii is dominated by background events consisting of charged particles that miss the FASER$\nu$ scintillator station. 
		}
\label{fig:sel_events} 	
\end{figure}

 \begin{table}[b]
    \begin{tabular}{llr}
      \toprule[1pt]
       Category & Events & Expectation   \\
       \midrule[0.75pt]
             Signal     & 153 &  $ n_{\nu} + n_b \cdot p_1 \cdot p_2 + n_{\mathrm{had}}  + n_{\mathrm{geo}} \cdot f_{\mathrm{geo}} $ \\ 
             $n_{10}$     & 4& $n_b \cdot (1-p_1) \cdot p_2$   \\ 
             $n_{01}$     & 6 & $n_b \cdot p_1 \cdot (1- p_2 )$\\ 
             $n_{2}$      & 64'014'695 & $n_b \cdot (1-p_1) \cdot (1-p_2) $ \\ 
      \bottomrule[1pt]
    \end{tabular}
  \caption{\label{tab:events} Observed event yields in \lumi of collision data and their relation to neutrino and background events.}
\end{table}

Figure~\ref{fig:sel_events} shows the selected events, as well as the background-enriched regions with lower momentum or $r_{\mathrm{veto}\, \nu} > $ \SI{120}{\milli\meter}. In total we observe 153 events passing all selection steps. Using \texttt{GENIE} we study the composition of neutrino events passing this selection and find that 99\% originate from muon neutrino CC interactions.

We group the selected events into four categories to estimate the number of neutrino ($n_\nu$) and background events ($n_b$). The categorization is determined by whether the events pass or fail the FASER$\nu$ veto scintillator selection criteria. This allows us to determine in a simulation-independent way the inefficiencies of the two layers of the FASER$\nu$ veto scintillator ($p_1$, $p_2$) under the assumption that they are uncorrelated. 


Besides the signal category, we select:
\begin{itemize}
\setlength\itemsep{-0.05in}
    \item[$n_{10}$:] Events for which the first layer of the FASER$\nu$ scintillator produces a charge of $>$\SI{40}{\pico\coulomb} in the PMT, but no signal with sufficient charge is seen in the second layer.
   \item[$n_{01}$:] Analogous events for which more than \SI{40}{\pico\coulomb} in the PMT was observed in the second layer, but not in the first layer.
     \item[$n_{2}$:] Events for which both layers observe more than \SI{40}{\pico\coulomb} of charge.
\end{itemize}
Table~\ref{tab:events} lists the observed event yields and their relation to the expected number of neutrino and background events and the FASER$\nu$ veto scintillator inefficiencies.

We analyze the observed number of events using a binned extended maximum likelihood fit, implemented using the \texttt{iminuit} package~\cite{iminuit}. We introduce nuisance parameters to constrain the estimated background events to their expectations using Gaussian priors.
The likelihood is numerically maximized, and we use a discovery test statistic~\cite{Cowan:2010js} to determine the significance of the observed signal over the background-only hypothesis. We find
\begin{align}
 n_\nu            &  = 153 \, {}^{+12}_{-13} \, (\mathrm{stat.}) \,\,{}^{+2}_{-2}  \, (\mathrm{bkg.}) = 153\,\, {}^{+12}_{-13} \, (\mathrm{tot.})  \, \nonumber
\end{align}
with a significance of \obssign\ standard deviations over the background-only hypothesis and based on the asymptotic distribution of the test statistic. The excess is compatible with the expected number of neutrino events \nuexp, but note that its error does not include any systematic uncertainties from simulating the detector response and selection. The determined inefficiencies of the two FASER$\nu$ scintillators are $ p_1 = ( 6^{+4}_{-3} ) \times 10^{-8}$ and $ p_2 = ( 9^{+4}_{-3} ) \times 10^{-8}$, showing values close to the expected performance~\cite{FASER:2022hcn}.

\begin{figure}[t]
	\centering
		 \includegraphics[width=1.0\linewidth]{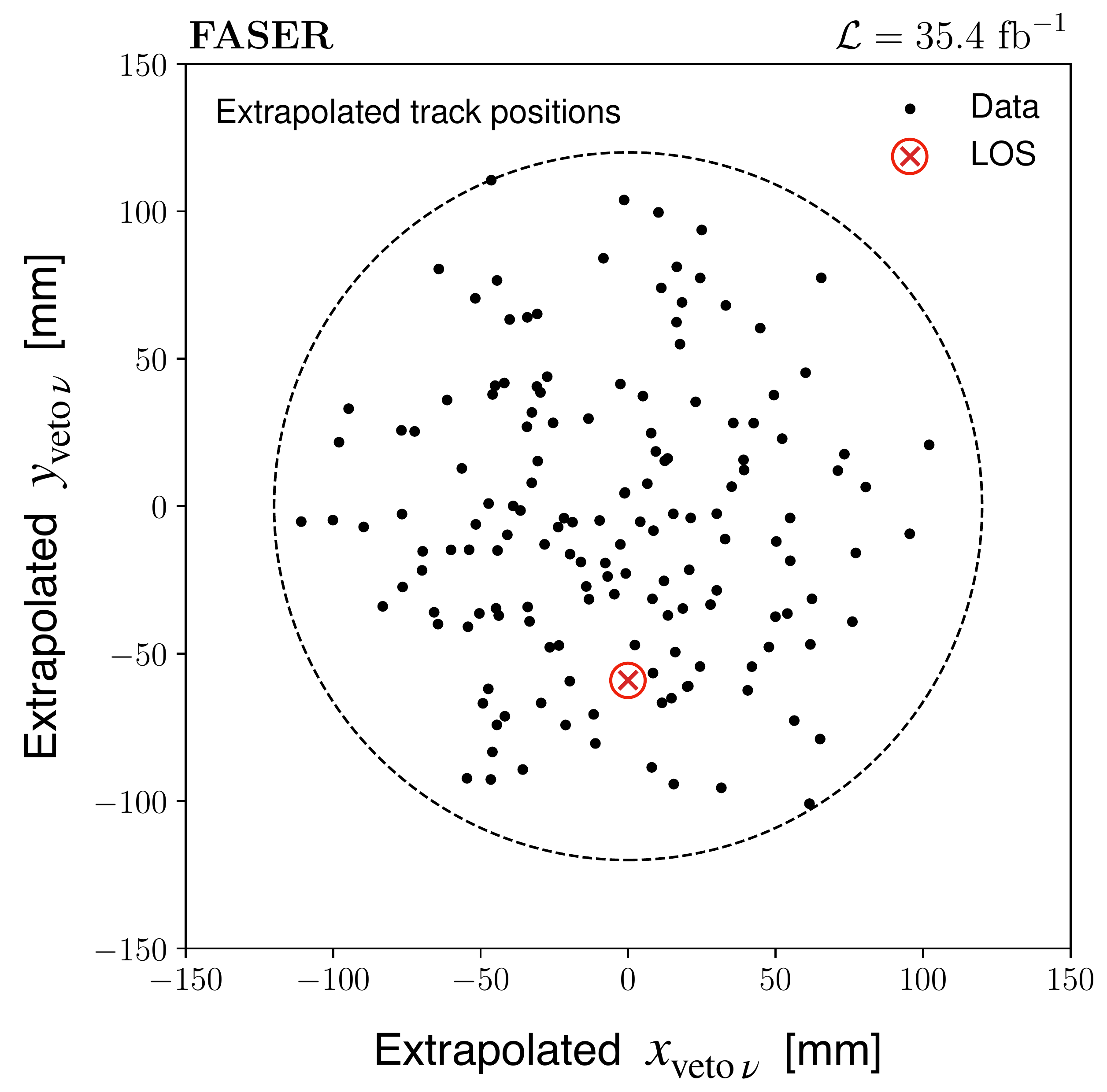}  
\vspace*{-0.1in}
   \caption{ 
	Extrapolated transverse position of the reconstructed tracks of neutrino-like events to the FASER$\nu$ scintillator station. The ATLAS LOS is indicated with a red marker and shifted \SI{59}{\milli\meter} in the negative $y$ direction from the center of the scintillator station. 
	}
\label{fig:FASERnu_scint} 	
\end{figure}

We expect that the identified neutrino candidates are distributed around the ATLAS LOS and do not cluster at a specific point of origin. We test this by using the extrapolated position to the FASER$\nu$ scintillator station from the reconstructed tracks of the neutrino-like events in the signal category. Figure~\ref{fig:FASERnu_scint} shows the extrapolated positions and we observe the expected behaviour. 

\begin{figure*}
	\centering
		 \includegraphics[width=0.24\linewidth]{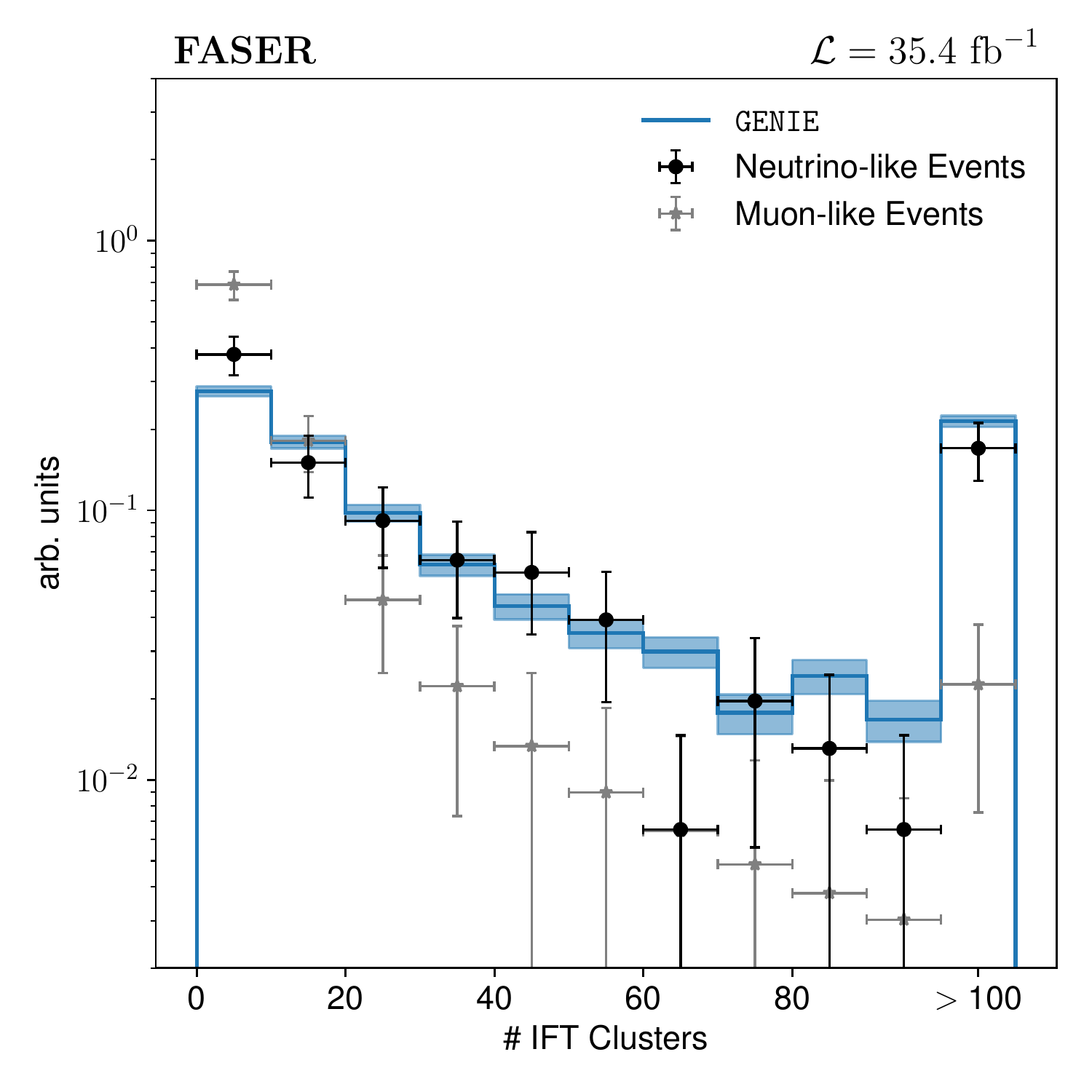}  		 		 
		  \includegraphics[width=0.24\linewidth]{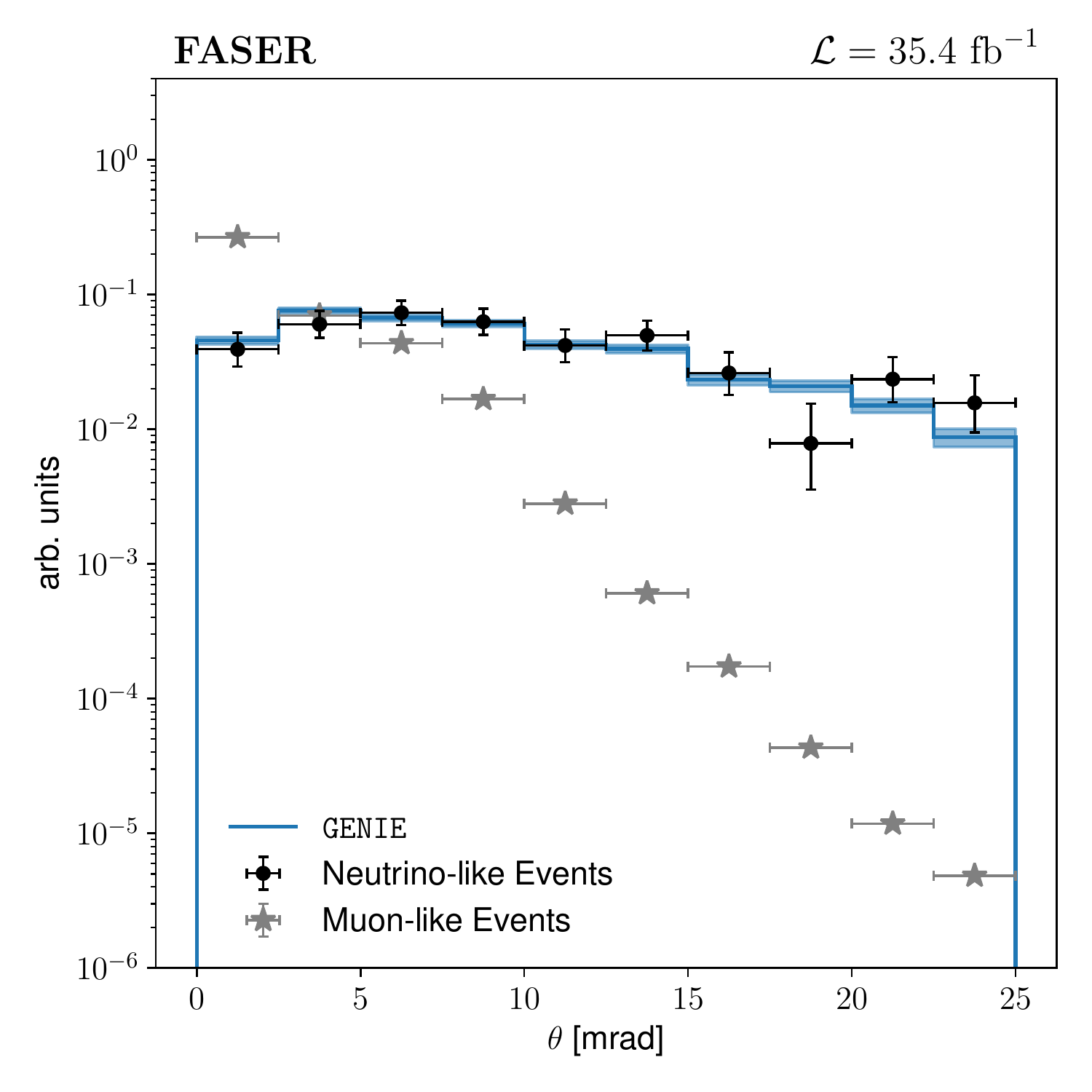}   
		  \includegraphics[width=0.24\linewidth]{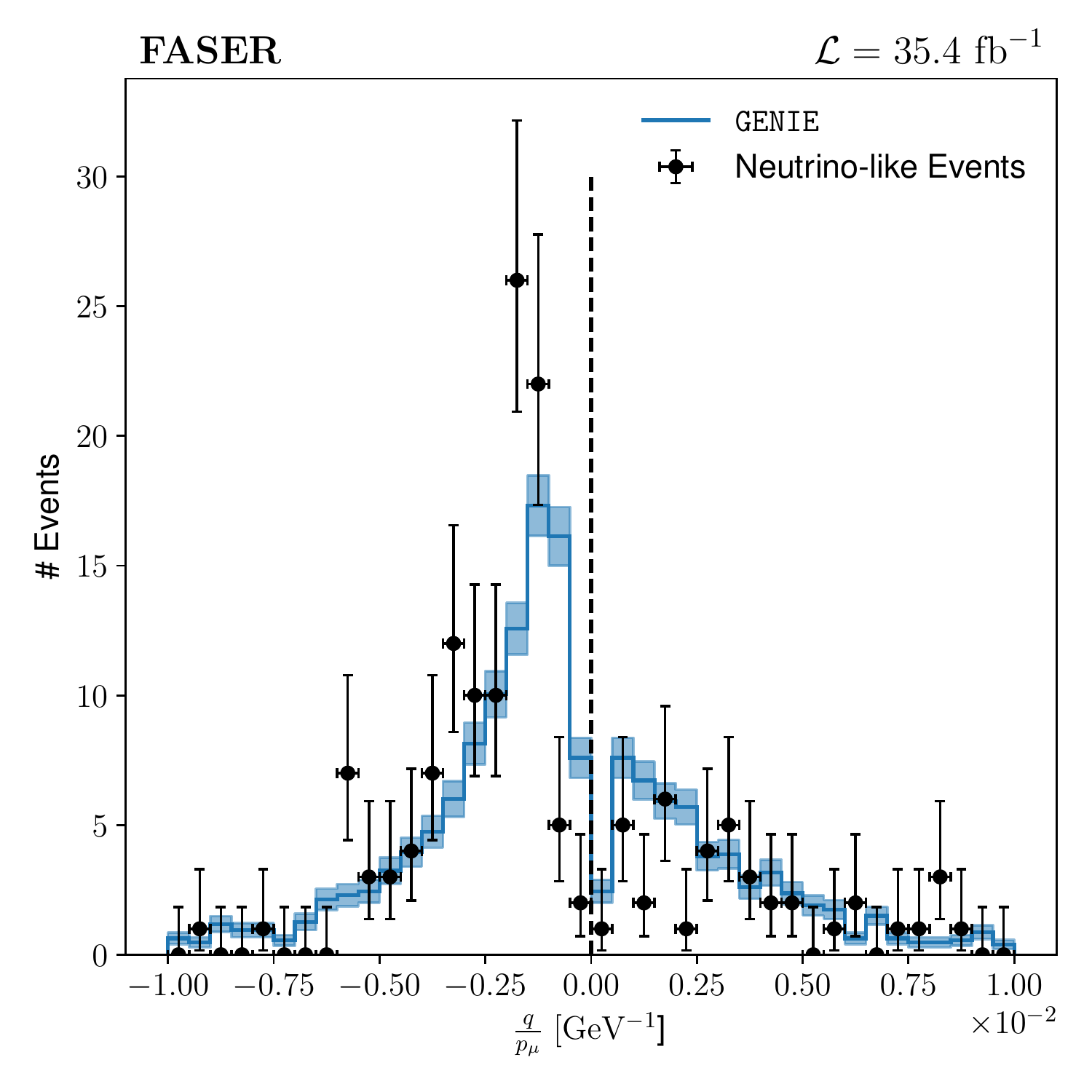}    
		  \includegraphics[width=0.24\linewidth]{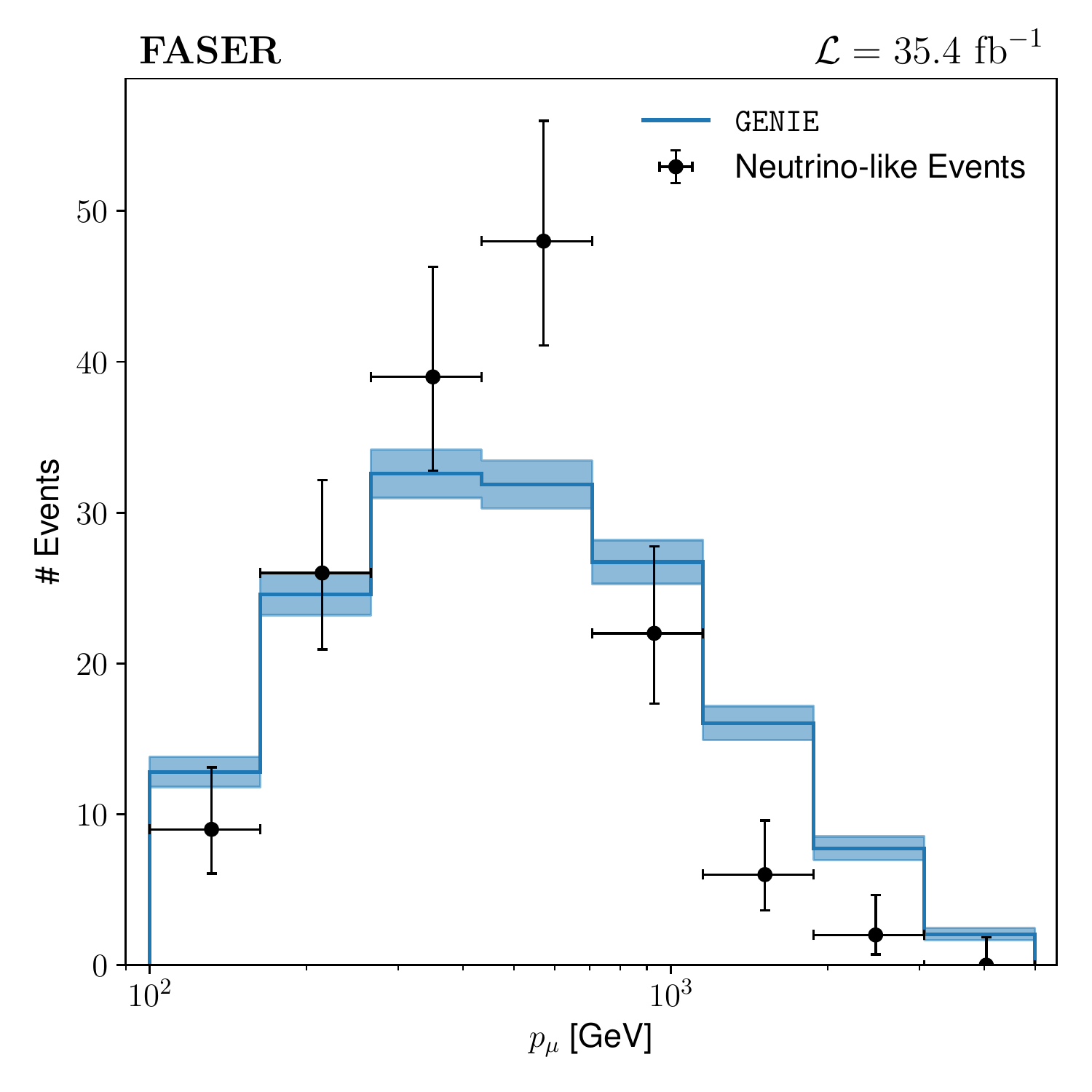}  
\vspace*{-0.1in}		  
   \caption{  The figures depict the number of reconstructed clusters in the IFT, track polar angle $\theta_\mu$,  $q/p_\mu$, and the reconstructed momentum $p_\mu$ for events in the signal region (black markers) and compare them to the expectation from \texttt{GENIE} (blue) and muon-like events (grey markers). The muon-like events are from the $n_2$ category, for which both layers of the FASER$\nu$ scintillator observed a signal, and show the expected distributions for non-neutrino backgrounds. The blue bands correspond to the statistical error of the simulated samples and are luminosity scaled for $q/p_\mu$ and $p_\mu$. The other figures are normalized to unity. 
	}
\label{fig:IFT_cluster_q_over_p_momentum} 	
\end{figure*}

Figure~\ref{fig:IFT_cluster_q_over_p_momentum} summarizes additional properties of the signal category events. The CC neutrino interactions produce on average a larger number of particles than MIP interactions, which appear in the IFT as charge depositions. The number of IFT clusters of the signal category is very distinct from background-like ($n_2$) events and agrees well with the expectation from \texttt{GENIE}. We also examine the polar angles $\theta_\mu$ of the neutrino candidates and observe distributions close to the simulated neutrino events and distinctively different from muon backgrounds. We observe a clear charge separation in $q/p_\mu$ for the reconstructed tracks, with $q$ denoting the assigned track charge. In total 40 events with a positively-charged track candidate are observed, showing the presence of anti-neutrinos in the analyzed data set. The reconstructed momentum of the muon produced in a CC $\nu_\mu$ interaction is a good proxy for the incident neutrino energy. Using the simulated CC neutrino interactions, we estimate that with our analysis strategy we select neutrino events for which on average $>80\%$ of the incident neutrino momentum is transferred to the final state muon. This indicates that a large fraction of the reconstructed neutrino candidates have energies significantly larger than \SI{200}{\GeV}. A detailed study of these properties, which accounts for systematic effects, is left for future work.

\papersection{Summary}

We report the first direct detection of neutrinos produced at a collider experiment using the active electronic components of the FASER detector. We observe $153^{+12}_{-13}$ neutrino events from CC interactions from $\nu_\mu$ and $\overline \nu_\mu$ taking place in the tungsten-emulsion detector of FASER$\nu$. The spatial distribution and properties of the observed signal events are consistent with neutrino interactions, and the chosen analysis strategy does not depend on the quality of the modeling of detector effects in the simulation. For the signal events, the reconstructed charge shows the presence of anti-neutrinos, and the reconstructed momentum implies that neutrino candidates have energies significantly above \SI{200}{\GeV}. This result marks the beginning of the field of collider neutrino physics, opening up a wealth of new measurements with broad implications across many physics domains~\cite{Feng:2022inv}.

\papersection{Acknowledgments}

We thank CERN for the very successful operation of the LHC during 2022. We thank the technical and administrative staff members at all FASER institutions for their contributions to the success of the FASER project. We thank the ATLAS Collaboration for providing us with accurate luminosity estimates for the used Run 3 LHC collision data. FASER gratefully acknowledges the donation of spare ATLAS SCT modules and spare LHCb calorimeter modules, without which the experiment would not have been possible. We also acknowledge the ATLAS collaboration software, Athena, on which FASER’s offline software system is based~\cite{ATL-PHYS-PUB-2009-011} and the ACTS tracking software framework~\cite{Ai:2021ghi}. Finally we thank the CERN STI group for providing detailed FLUKA simulations of the muon fluence along the LOS, which have been used in this analysis. This work was supported in part by Heising-Simons Foundation Grant Nos.~2018-1135, 2019-1179, and 2020-1840, Simons Foundation Grant No.~623683, U.S.~National Science Foundation Grant Nos.~PHY-2111427, PHY-2110929, and PHY-2110648, JSPS KAKENHI Grants Nos.~JP19H01909, JP20K23373, JP20H01919, JP20K04004, and JP21H00082, BMBF Grant No.~05H20PDRC1, DFG EXC 2121 Quantum Universe Grant No. 390833306, ERC Consolidator Grant No.~101002690, Royal Society Grant No.~URF\textbackslash R1\textbackslash 201519, UK Science and Technology Funding Councils Grant No. ST/ T505870/1, the National Natural Science Foundation of China, Tsinghua University Initiative Scientific Research Program, and the Swiss National Science Foundation. 

\bibliographystyle{utphys}
\bibliography{faser}

\clearpage
\newpage
\onecolumngrid
\section{Appendix} 

\subsection{Appendix~A: Geometric Sideband}\label{app:a}

\begin{figure}[h!]
	\centering
		 \includegraphics[width=0.45\linewidth]{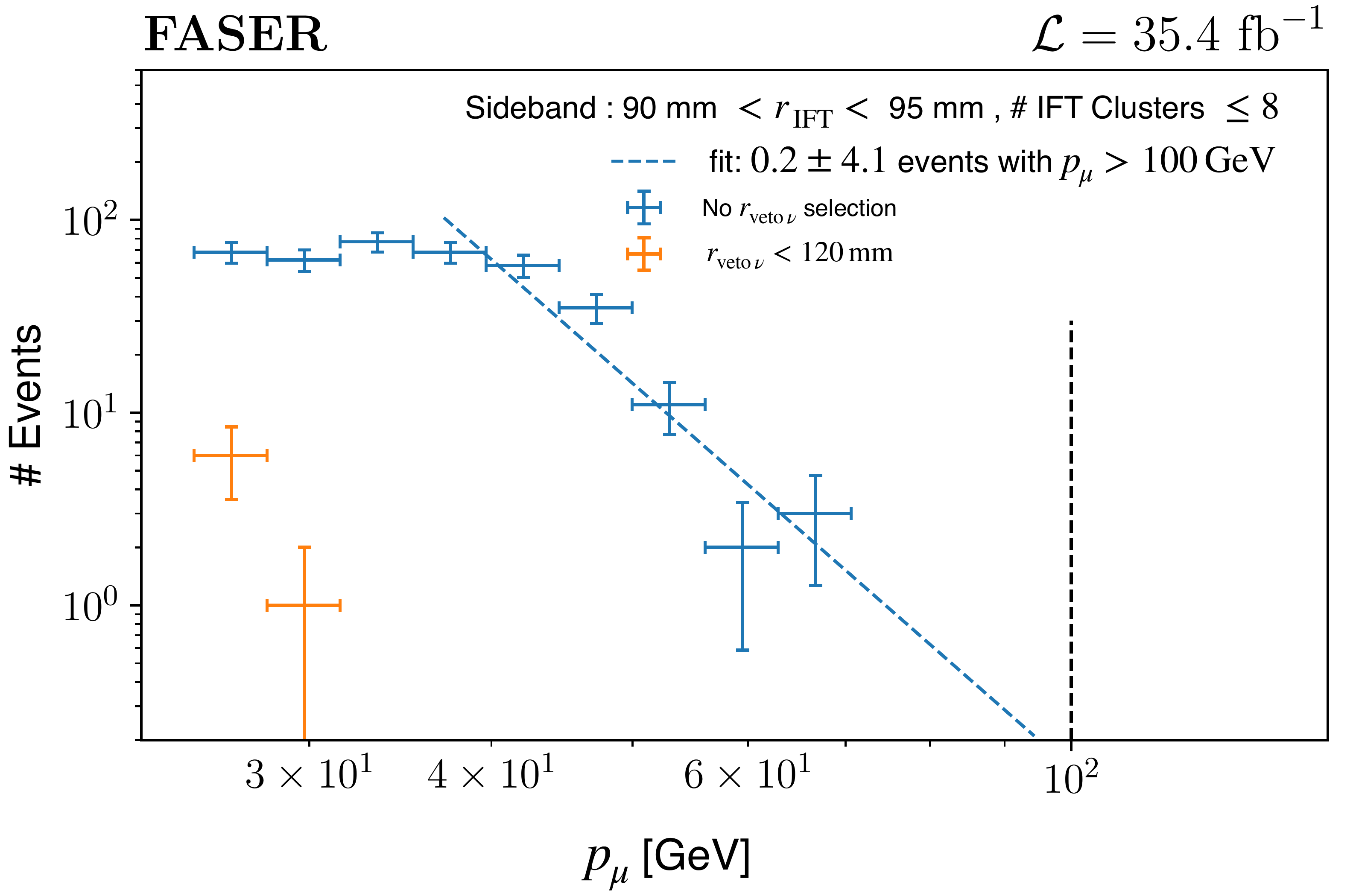}  
\vspace*{-0.1in}
   \caption{ 
	Sideband for geometric background estimation. 
	}
\label{fig:Signal_region} 	
\end{figure}

Figure~\ref{fig:Signal_region} depicts the sideband used to estimate the geometric backgrounds of the analysis. Background events are required to be consistent with a muon candidate by having $\leq 8$ IFT clusters and an extrapolated radius $r_{\mathrm{IFT}}$ of \SI{90}{\milli\meter} to \SI{95}{\milli\meter} with respect to the IFT center. This selection is dominated by geometric background events that do not pass the signal selection steps of the analysis. No events with $p_\mu > \SI{100}{GeV}$ are observed. To estimate the number of events within this momentum range, we linearly extrapolate the events between \SI{30}{GeV} and \SI{100}{GeV}  and find $0.2 \pm 4.1$ events, with the error denoting the statistical error. To account for the $r_{\mathrm{veto}\,\nu}$ requirement of the signal selection, we further apply a requirement of  $r_{\mathrm{veto}\, \nu} < 120 \, \mathrm{mm}$ to the sideband events (orange distribution). No events with $p_\mu > 30 \, \mathrm{GeV}$ are observed. We thus use \num{5.9} as the 3$\sigma$ upper limit and use this to calculate the ratio with respect to the number of events without any $r_{\mathrm{veto}\, \nu}$ selection, to correct the sideband background events for the $r_{\mathrm{veto}\, \nu}$ requirement. With this factor we find $n_{\mathrm{geo}} = 0.01 \pm 0.23$ geometric background events. To account for the fact that this number corresponds to an annulus, the correction factor $f_{\mathrm{geo}} = 7.9 \pm 2.4$, determined from simulation, is applied. It is obtained from simulation with the uncertainty spanning different assumptions about the angle, momenta, and positions of the geometric background events. 

\subsection{Appendix~B: Event Display}\label{app:b}
\ 

\begin{figure}[htbp]
	\centering
		 \includegraphics[width=0.99\linewidth]{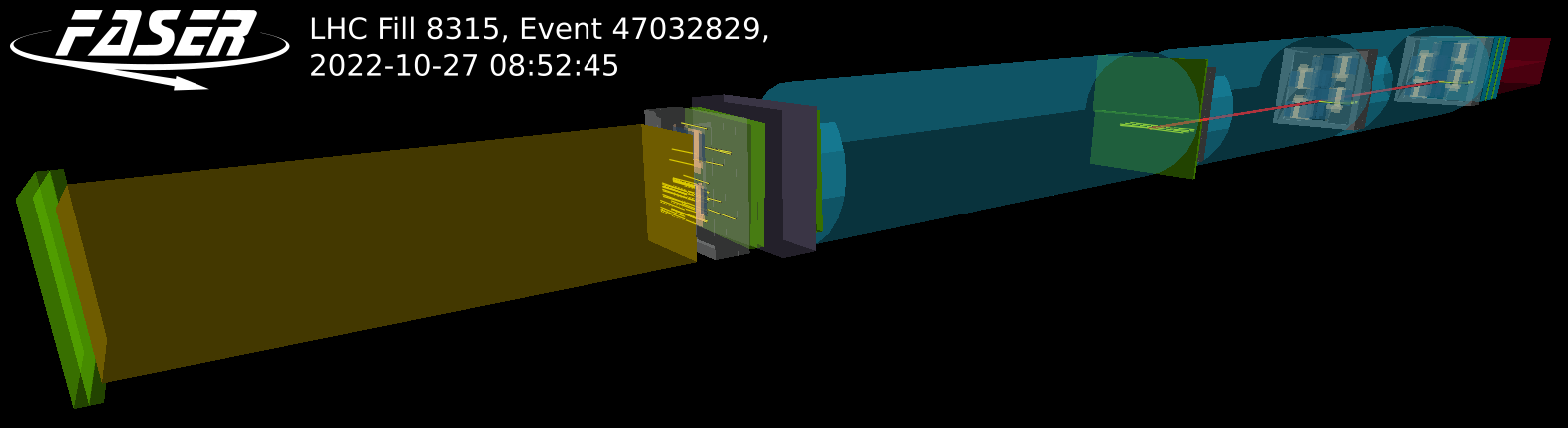}  
\vspace*{-0.1in}
   \caption{ 
	 Event display of a neutrino interaction candidate in which secondary particles produced in the CC interaction produce activity in the IFT. 
	}
\label{fig:evtdisplay} 	
\end{figure}

Figure~\ref{fig:evtdisplay} shows an event display of an example neutrino candidate event. The event has a momentum of $p_\mu = \SI{843.9}{\GeV}$, negative charge, 
$\theta_\mu = 2.5 \,\text{mrad}$, $r_{\mathrm{veto}\, \nu}  = \SI{57.2}{\milli\meter}$, $r_{\mathrm{IFT}} = \SI{55.8}{\milli\meter}$ and produced 57 clusters in the IFT. 

\subsection{Appendix~C: Likelihood Fit}\label{app:c}

\begin{figure}[htbp]
	\centering
		 \includegraphics[width=0.4\linewidth]{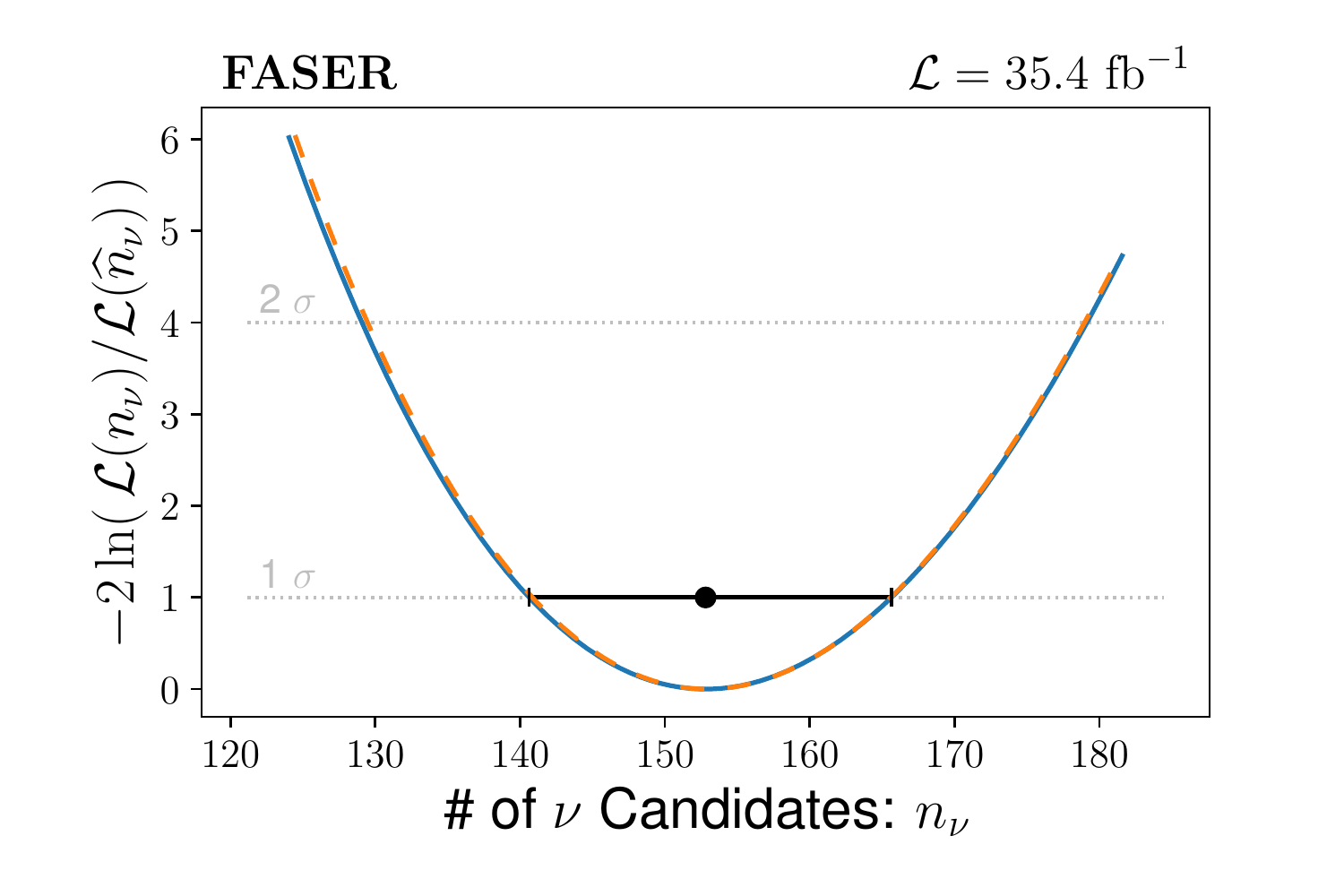}
\vspace*{-0.1in}
   \caption{ 
	The log-likelihood ratio of the estimated number of neutrinos is shown in blue. The dashed orange contour fixes the parameters of $\mathcal{G}_i$ to determine the statistical uncertainty of the neutrino signal yield.
	}
\label{fig:NLL_p_nu} 	
\end{figure}

The used likelihood has the form
\begin{align}
 \mathcal{L} = \prod_{i} \mathcal{P}( N_i | n_i ) \cdot \prod_j \mathcal{G}_j  \, .
\end{align}
Here $\mathcal{P}$ denotes a Poissonian with the index $i$ running over the four event categories with observed event counts $N_i$ and expectation values $n_i$. We introduce nuisance parameters to constrain the estimated number of background events to their expectations using three Gaussian priors $\mathcal{G}_j$ . The used test statistic has the form
\begin{align}
  q_0 = \bigg\{ \begin{matrix} - 2 \ln \lambda(n_\nu = 0) & \widehat n_\nu \ge 0 \\ 0 & \widehat n_\nu < 0  \end{matrix}
\end{align}
and the significance of the observed signal $\widehat n_\nu$ over the background-only hypothesis is given by $\sqrt{q_0}$ in the asymptotic limit. Further \mbox{$\lambda(n_\nu = 0) := \mathcal{L}(n_\nu) / \mathcal{L}(\widehat n_\nu)$} denotes the ratio of the likelihood maximized with the condition of no signal, $n_\nu = 0$, to the unconditionally maximized likelihood. The log-likelihood ratio is shown in Figure~\ref{fig:NLL_p_nu}.

\vspace{-4ex}

\subsection{Appendix~D: Momentum Resolution}\label{app:d}

Data-driven alignment corrections are applied to the positions and orientations of the modules of the tracking spectrometer stations using a sample of reconstructed muons. In the case of perfect alignment of the FASER tracking detectors, we expect a momentum resolution of 2.1\% at 100 GeV, 4.7\% at 300 GeV, and 16.4\% at 1 TeV. The accuracy of the alignment is validated using a photon conversion sample for momenta up to 250 GeV.


\subsection{Appendix~E: Expected Number of Neutrino Events}\label{app:e}

\begingroup

\renewcommand{\arraystretch}{1.2}

 \begin{table}[b]
    \begin{tabular}{l|c|ccc|c|c}
	\toprule[1pt]
       Volume & Type & $0 < E_\nu < \SI{500}{\GeV}$ & $500 < E_\nu < \SI{1000}{\GeV}$ & $E_\nu > \SI{1000}{\GeV}$  & $\sum$ & $\overline E_\nu$ [GeV]  \\
	\midrule[0.75pt]
             FASER$\nu$     & $\nu_\mu$ &359 / 379 & 239 / 273 & 291 / 790 & 890 / 1442 & 880 / 1376  \\
             FASER$\nu$     & $\overline \nu_\mu$ &  116 / 130 & 62 / 85 & 49 / 151 & 227 / 367 & 657 / 1028 \\
             	\midrule[0.75pt]
             $r < \SI{95}{\milli\meter}$    & $\nu_\mu$ & 147 / 154 & 105 / 118 & 141 / 375 & 394 / 647 & 943 / 1477  \\ 
            $r < \SI{95}{\milli\meter}$    & $\overline \nu_\mu$ & 48 / 53 & 28 / 37 & 23 / 67 & 99 / 157 & 687 / 1057 \\
   	\bottomrule[1pt]
    \end{tabular}
    \vspace*{-0.05in}
  \caption{\label{tab:preds} 
  The expected numbers of neutrino and anti-neutrino events from \texttt{SIBYLL} (first number) and \texttt{DPMJET} (second number) for an integrated luminosity of \lumi\ i and different energy intervals, along with the sum over all energy intervals, and the average neutrino energy $\overline{E}_{\nu}$.  Results are shown requiring the interactions to be (1) in the FASER$\nu$ detector volume or (2) in the target region and within a radius of \SI{95}{\milli\meter} from the center of the FASER detector.   }
\end{table}

\endgroup

The predicted numbers of neutrino and anti-neutrino interactions from \texttt{SIBYLL} and \texttt{DPMJET} are listed in Table~\ref{tab:preds}. Results are shown requiring the interactions to be (1) in the FASER$\nu$ detector volume or (2) in the target region and within a radius of \SI{95}{\milli\meter} from the center of the FASER detector. Note that no additional acceptance and efficiency corrections are applied and the second requirement approximates the fiducial volume used in the analysis.

\end{document}